\documentclass[a4paper,11pt]{article}
\usepackage{pos}
\usepackage{natbib}
\usepackage{subfigure}
\usepackage{comment}
\usepackage{ulem}

\hypersetup{citecolor=blue}

\title{TeV emission from FSRQs: the first systematic and unbiased survey}
 \ShortTitle{Search for TeV emission from FSRQs}

\author*[a]{Patel, S. R.}

\affiliation[a]{Deutsches Elektronen-Synchrotron DESY,\\
  Platanenallee 6, D-15738 Zeuthen, Germany}

\forColl{VERITAS} 

\emailAdd{sonal.patel@desy.de}

\abstract{Flat spectrum radio quasars (FSRQs) have been detected at TeV energies by ground-based atmospheric Cherenkov telescopes mainly during flaring states. VERITAS is carrying out the first systematic and unbiased search for TeV emissions from a set of FSRQs. Fermi-LAT-detected FSRQs with positive declinations and extrapolated fluxes from the 3FHL catalog exceeding 1$\%$ Crab at >200 GeV after correcting for EBL absorption were selected for this survey, resulting in eight targets. Additionally, four FSRQs that were already detected at TeV energies are also included in this survey. In an unbiased fashion, the observations of 12 FSRQs, even without detection, will provide the first constraints on their duty cycle of TeV emission. We report the results from four of the 12 FSRQs observed during 2020-21 season in this work. For these sources, we also show the results from nearly simultaneous \textit{Fermi}-LAT observations.

}

\FullConference{37$^{\rm{th}}$ International Cosmic Ray Conference (ICRC 2021)\\
		July 12th -- 23rd, 2021\\
		Online -- Berlin, Germany}

\begin{document}
\maketitle

\section{Introduction}

Blazars, an ever-surprising class of active galactic nuclei, show two broad humps in the broadband spectral energy distributions (SEDs). The first low-frequency hump is attributed to synchrotron emission  from  relativistic  electrons,  gyrating  in  the  magnetic field of the jet.  Blazars can be classified depending on the location of the synchrotron emission peak in the SEDs.  The synchrotron emission peak of BL Lacs lies within the infrared-to-X-ray frequency range, while that of  FSRQs lies 
mostly within the infrared-optical frequency range. The  origin of  the  higher  frequency  hump  in  SEDs  is usually considered as inverse  Compton  (IC)  scattering  of  relativistic  electrons on the synchrotron photons (Synchrotron Self Compton, SSC) or the photons external to the jet (External Compton, EC). The external photon field for Comptonization are mostly associated with the  accretion  disk \citep{Dermer1993}, broad-line region (BLR) or dusty torus (DT) \citep{Sikora1994, Blazejowski2000}.  The SSC emission is generally used to explain the high-frequency hump in BL Lacs, while EC emission is preferred for FSRQs due to the presence of seed photons in the circumnuclear environment of these sources, evident from characteristic emission lines in their optical spectra. Alternatively, in the hadronic scenario, it is possible to produce high energy $\gamma$-ray via $p\gamma$ interaction followed by proton synchrotron or neutral pion decay processes \citep{Mannheim1992, Mastichiadis1995, Mucke2001, Bottcher2013}. Recently, it has been shown that FSRQs are expected to be sources of astrophysical neutrinos in the sub-EeV range \cite{Righi2020}.


Generally FSRQs are located at larger distances compared to BL Lacs \citep{Ghisellini2010}. The attenuation of very high energy $\gamma$-ray from these distant objects due to scattering on the extra-galactic background light and possible intrinsic spectral cutoff above GeV energies \citep{Ghisellini1998} make their detection difficult by the current generation of ground-based imaging atmospheric Cherenkov telescopes (IACTs), namely the High Energy Stereoscopic System \citep[H.E.S.S.,][]{Trichard2015}, Major Atmospheric Gamma Imaging Cherenkov Telescope \citep[MAGIC,][]{MAGIC2016} and Very Energetic Radiation Imaging Telescope Array System \citep[VERITAS,][]{Park2015}. So far a total of eight FSRQs are detected at TeV energies by these three major ground-based $\gamma$-ray detectors and are listed in Table~\ref{tab:ListFSRQs}. All of them were detected during high flux states, except PKS 1510-089, from which TeV emission has also been seen in its low flux state \citep{Acciari2018}. 

\begin{table}[!ht]
\centering
\caption{FSRQs detected by major IACTs.}
\label{tab:ListFSRQs}
\begin{tabular}{lccc}
\hline
Source Name  & $z$ & Detected by & Reference \\
\hline
PKS 1222+216 & 0.43 & MAGIC, VERITAS & \citep{Mose2010, Holder2014}\\
Ton~599      & 0.73 & MAGIC, VERITAS & \citep{Mirzoyan2017Ton599, Mukherjee2017}\\
PKS 1441+25  & 0.94 & MAGIC, VERITAS & \citep{Mirzoyan2015, Mukherjee2015}\\
3C~279       & 0.54 & H.E.S.S., MAGIC &  \citep{MAGIC2008, Cerruti2017} \\
PKS~1510-089 & 0.36 & H.E.S.S., MAGIC & \citep{HESS2013, Cortina2012} \\
B2 1420+32  & 0.68 & MAGIC & \citep{Mirzoyan2020} \\
S3 0218+357 & 0.94 &  MAGIC & \citep{Mirzoyan2014} \\
PKS 0736+017 & 0.19 & H.E.S.S. & \cite{HESS2020} \\
\hline     
\end{tabular}
\end{table}



In this work, we present the results from VERITAS observations of four FSRQs: GB6 J0043+342, S3 0218+35, PKS 0736+017 and 3C 454.3. These observations are part of the survey of TeV emission from FSRQs, which is discussed in section ~\ref{sec:survey}.  The details of observation and data analysis are mentioned in section~\ref{sec:data}. We then discuss our results in section~\ref{sec:result}.


\section{VERITAS survey of TeV emission from FSRQs}
\label{sec:survey}

\begin{figure}[!ht]
\subfigure[Sources included in the survey are shown on galactic coordinate system. This work presents the results from the sources shown in red text. \label{fig:sources}]{\includegraphics[scale=0.38]{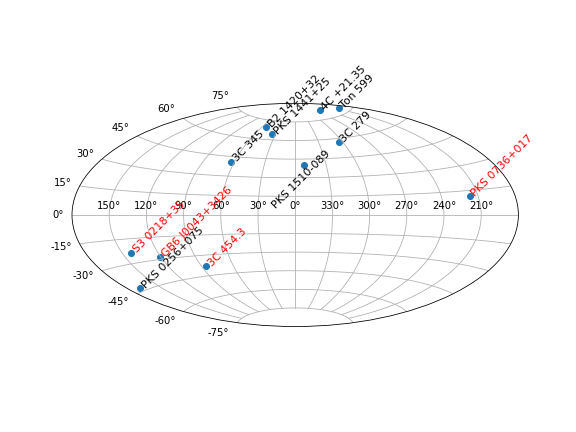}}
\hspace{0.15cm}
\subfigure[VERITAS observations of four sources during 2020-21 season. The legends correspond to the starting MJD on the horizontal axis for each source.\label{fig:observation}]{\includegraphics[scale=0.45]{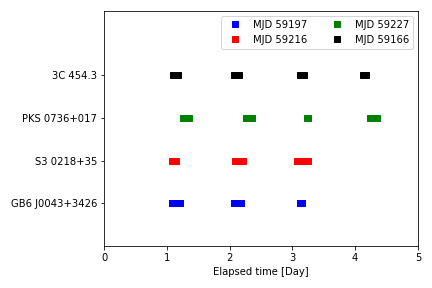}}
\caption{VERITAS survey of TeV emission from FSRQs}
\end{figure}

We have selected 12 FSRQs showing relatively hard GeV spectra in this survey, as shown in Figure~\ref{fig:sources}. The sample includes eight sources with positive  declination  having an  extrapolated 3FHL catalog flux in excess of 1$\%$ of the Crab at 200 GeV after taking into account EBL absorption \citep{Dominguez2011}. The remaining four FSRQs were already detected at TeV energies: PKS 0736+017 and B2 1420+32 have northern declination but extrapolated fluxes $<$1$\%$ Crab, and 3C 279 and PKS 1510-089 are among the most active TeV blazars but culminate below 60$^\circ$elevation for VERITAS. 

For each of these sources, eight hours of observation split over two consecutive nights, if possible, were requested. Our goal was to achieve the deepest possible VERITAS instantaneous sensitivity without compromising the energy threshold  by  going  to  too  low an elevation. 
Four of these sources were observed during the season of 2020-21, with observations spanning  three nights for GB6 J0043+3426 and S3 0218+35, and four nights for PKS 0736+017 and 3C 454.3, as shown in Figure~\ref{fig:observation}.

The data set of this survey aims to provide the first unbiased characterisations of TeV emission from the jets of FSRQs. A subset of these observations that yields  non-detection will constrain the  extrapolation  of  the  LAT  fluxes for these sources, as shown in the present work. The quantification of possible  spectral  breaks  is necessary  to accommodate the VERITAS upper limits, which will provide valuable constraints to the dominant cooling mechanism of the highest-energy electrons in the emitting plasma, and could identify the energy of the seed photon field for Compton up-scattering. Additional value of the resulting data set comes from the systematic acquisition of TeV data on a  small  but  representative  population  of  FSRQs.   Even  if  no  clear  detection  comes  out  of  this study, it would already provide the first quantitative statement on the very limited duty cycle of TeV emission from the population of FSRQs.

\section{Observation and Data analysis}
\label{sec:data}

VERITAS is one of the most sensitive ground-based $\gamma$-ray detectors. It is located at the Fred Lawrence Whipple Observatory in southern Arizona (31 40N, 110 57W, 1.3km a.s.l.). It is sensitive to $\gamma$-ray emission in the energy range, 100 GeV to $>$30 TeV. The VERITAS array has four 12 m diameter, 12 m focal length imaging atmospheric Cherenkov telescopes. Each telescope has a Davies-Cotton design segmented mirror dish of 345 facets and focuses the Cherenkov light from particle showers onto a pixellated camera having 499 PMTs and a total field of view of 3.5$^\circ$. The current configuration of the array can detect an object having 1$\%$ Crab Nebula flux in $\sim$25 hours with energy resolution of 15-25$\%$. For a 1 TeV photon, the 68$\%$ containment radius is $<0.1^\circ$, with a pointing accuracy of $<$50". As a part of the blazar monitoring program, VERITAS has collected $\sim$30 hr of good quality data from four FSRQs during 2020-21. The VERITAS software EventDisplay \citep{Krause2017,  Maier2017} based on image parameters and boosted decision trees for gamma hadron separation was used to analyse these data. Data were also analysed and compared with a second independent VERITAS software.

\begin{figure}[!ht]
\centering
\subfigure[GB6 J0043+3426]{\includegraphics[scale=0.4]{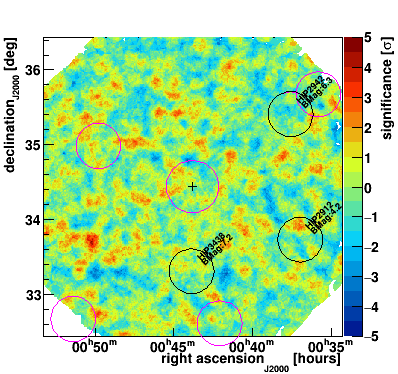}}
\subfigure[S3 0218+35]{\includegraphics[scale=0.4]{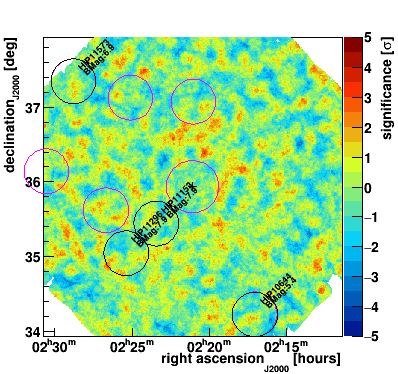}}

\subfigure[PKS 0736+017]{\includegraphics[scale=0.4]{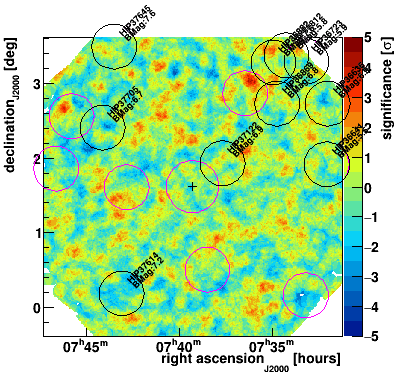}}
\subfigure[3C 454.3]{\includegraphics[scale=0.4]{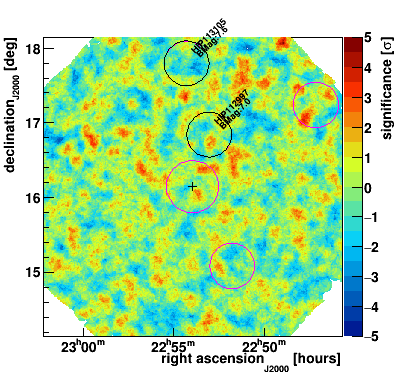}}
\caption{Significance maps of four FSRQs studied in this work. The circles indicate the regions excluded for background estimation. The excluded regions are the source region and stars brighter than eighth magnitudes.}
\label{fig:SigMap}
\end{figure}


The high energy data in the energy range from 0.1-300 GeV were obtained from the Large Area Telescope on board \textit{Fermi} satellite \citep[\textit{Fermi}-LAT,][]{Atwood2009} which is a pair conversion $\gamma$-ray telescope orbiting in space since 2008 August. Data over a period of two months centered on the VERITAS observations were used in this work. These data were analysed using \texttt{Fermitools$-$1.2.23} and \texttt{Fermipy-0.17.4} \citep{Wood2017}.  For all four sources, the events were extracted from the region of interest (ROI) of 15$^\circ$ centred around source position.  To avoid the  contamination  of  background $\gamma$ rays  from  Earth’s limb, a zenith  angle  cut  of  90$^\circ$ was  applied. To select the good time interval a filter of ‘(DATAQUAL$>$0)$\&\&$(LATCONFIG==1)’ was applied. The likelihood analysis \citep{Cash1979, Mattox1996} was performed, along with the galactic diffuse emission and the isotropic  background  models, and post-launch instrument response function (P8R3$\_$SOURCE$\_$V2v1). The models used for these four sources were as given in the 4FGL catalog \citep{Abdollahi2020} -- a log parabola for GB6 J0043+3426, S3 0218+35 and PKS 0736+017, and a power law with exponential cut off for 3C 454.3.

\begin{figure*}[!ht]
\centering
\includegraphics[scale=0.6]{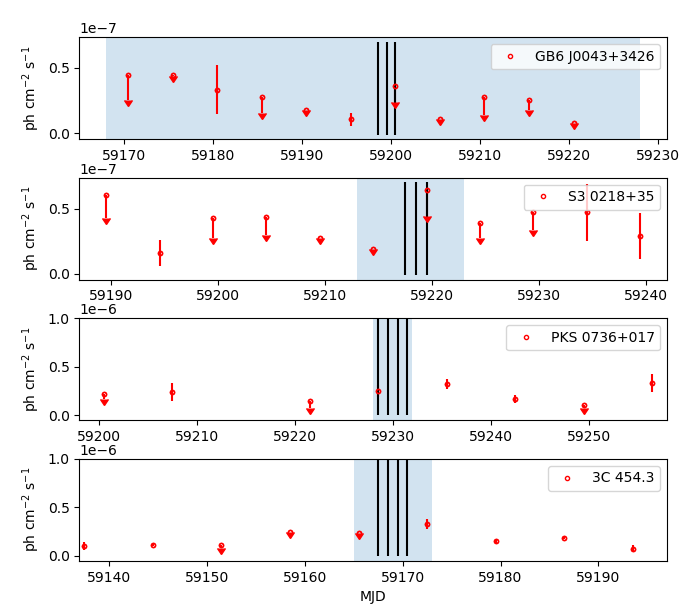}
\caption{0.1-300 GeV \textit{Fermi}-LAT light curves around VERITAS observations shown in black lines. The shaded regions correspond to the period of time-averaged LAT SED, shown in \ref{fig:spectra}. The test statistics of the averaged flux over shaded regions are, $\sim$17, $\sim$25, $\sim$34, and $\sim$58, for GB6 J0043+3426, S3 0218+35, PKS 0736+017, and 3C 454.3, respectively.}
\label{fig:lc}
\end{figure*}

\section{Results}
\label{sec:result}

Results from the VERITAS analysis are shown in Table~\ref{tab:results} and significance sky maps are shown in Figure~\ref{fig:SigMap}, for all four sources. It can be seen that none of these sources were detected during the observations carried out in 2020-21 season.  We compute the 95$\%$ flux upper limit (UL) assuming a power law model with spectral index of -3.5. Since VERITAS' operating energy range overlaps the high end of the \textit{Fermi}-LAT energy band, we also analysed the high energy \textit{Fermi}-LAT data. To have \textit{Fermi}-LAT SEDs as simultaneous as possible, we choose the periods around the VERITAS observations such that test statistics (TS) over 0.1-300 GeV is $\gtrsim$ 25. These periods are shown in Figure~\ref{fig:lc}.  All sources met the chosen criteria, except GB6 J0043+3426, for which data averaged over the period of two months did not pass this TS thresholds.

\begin{table}[!ht]
\centering
\caption{Analysis results for the season 2020-21}
\resizebox{\columnwidth}{!}{
\begin{tabular}{llcccccc}
\hline
Source &  Observation  & Live time & Significance &  Energy threshold & Flux (above E$_{th}$) \\
       &  period [MJD] &  [min]     &  $\sigma$    & E$_{th}$ [GeV]  & 10$^{-12}$ cm$^{-2}$ s$^{-1}$ ($\%$ Crab) \\
\hline
GB6 J0043+3426 & 59198-59201 & $\sim$425 & 0.53 & $\sim$170 & <5.74  ($\sim$1.6)\\ 
S3 0218+35     & 59217-59220 & $\sim$438 & -0.09 & $\sim$170 & <3.66 ($\sim$1.0)\\
PKS 0736+017   & 59228-59232 & $\sim$564 & -0.92 & $\sim$240 & <0.97 ($\sim$0.27)\\
3C 454.3       & 59167-59171 & $\sim$367 & -1.28 & $\sim$180 & <1.26 ($\sim$0.35)\\
\hline
\end{tabular}
\label{tab:results}
}
\end{table}


\begin{figure}[!ht]
\centering
\includegraphics[scale=0.45]{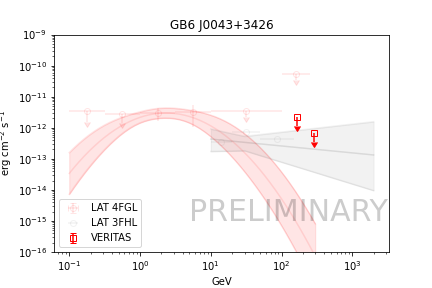}
\includegraphics[scale=0.45]{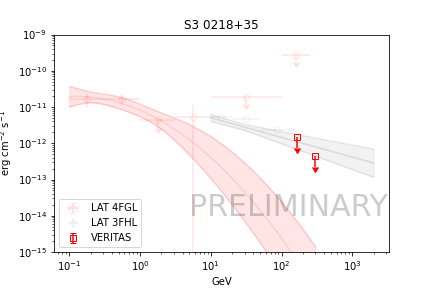}
\includegraphics[scale=0.45]{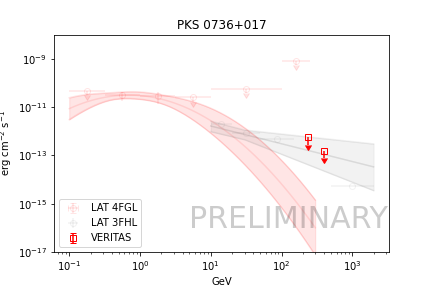}
\includegraphics[scale=0.45]{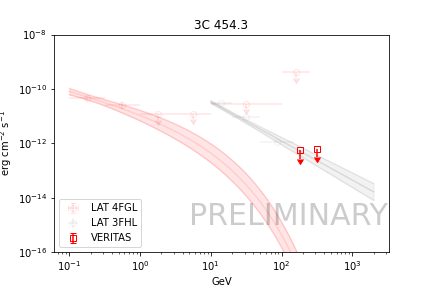}
\caption{VERITAS upper limits at energy threshold and decorrelation energy.}
\label{fig:spectra}
\end{figure}

Figure~\ref{fig:spectra} shows the observed SEDs of the four sources, with the VERITAS 95$\%$ UL at the energy threshold and decorrelation energy. The power law spectra from 10 GeV to 2 TeV are also shown from the 3FHL catalog \citep{Ajello2017}. The spectral indices as given in the catalog are 2.22$\pm$0.48, 2.54$\pm$0.17, 2.74$\pm$0.43 and 3.44$\pm$0.13, for GB6 J0043+3426, S3 0218+35, PKS 0736+017, and 3C 454.3, respectively. There was no significant detection of $\gamma$-ray from any of these  four sources above VERITAS energy threshold. However, the additional observations of these sources and remaining sources from the sample in the next season will help in providing stringent constrains on energy spectra above $\sim$200 GeV, where VERITAS has better point source sensitivity than the \textit{Fermi}-LAT. 




\section*{Acknowledgements}

This research is supported by grants from the U.S. Department of Energy Office of Science, the U.S. National Science Foundation and the Smithsonian Institution, by NSERC in Canada, and by the Helmholtz Association in Germany. This research used resources provided by the Open Science Grid, which is supported by the National Science Foundation and the U.S. Department of Energy's Office of Science, and resources of the National Energy Research Scientific Computing Center (NERSC), a U.S. Department of Energy Office of Science User Facility operated under Contract No. DE-AC02-05CH11231. We acknowledge the excellent work of the technical support staff at the Fred Lawrence Whipple Observatory and at the collaborating institutions in the construction and operation of the instrument.

\bibliography{BlazarPaper}
\bibliographystyle{aasjournal}

%
%
%

\clearpage \section*{Full Authors List: \Coll\ Collaboration}

\scriptsize
\noindent
C.~B.~Adams$^{1}$,
A.~Archer$^{2}$,
W.~Benbow$^{3}$,
A.~Brill$^{1}$,
J.~H.~Buckley$^{4}$,
M.~Capasso$^{5}$,
J.~L.~Christiansen$^{6}$,
A.~J.~Chromey$^{7}$, 
M.~Errando$^{4}$,
A.~Falcone$^{8}$,
K.~A.~Farrell$^{9}$,
Q.~Feng$^{5}$,
G.~M.~Foote$^{10}$,
L.~Fortson$^{11}$,
A.~Furniss$^{12}$,
A.~Gent$^{13}$,
G.~H.~Gillanders$^{14}$,
C.~Giuri$^{15}$,
O.~Gueta$^{15}$,
D.~Hanna$^{16}$,
O.~Hervet$^{17}$,
J.~Holder$^{10}$,
B.~Hona$^{18}$,
T.~B.~Humensky$^{1}$,
W.~Jin$^{19}$,
P.~Kaaret$^{20}$,
M.~Kertzman$^{2}$,
T.~K.~Kleiner$^{15}$,
S.~Kumar$^{16}$,
M.~J.~Lang$^{14}$,
M.~Lundy$^{16}$,
G.~Maier$^{15}$,
C.~E~McGrath$^{9}$,
P.~Moriarty$^{14}$,
R.~Mukherjee$^{5}$,
D.~Nieto$^{21}$,
M.~Nievas-Rosillo$^{15}$,
S.~O'Brien$^{16}$,
R.~A.~Ong$^{22}$,
A.~N.~Otte$^{13}$,
S.~R. Patel$^{15}$,
S.~Patel$^{20}$,
K.~Pfrang$^{15}$,
M.~Pohl$^{23,15}$,
R.~R.~Prado$^{15}$,
E.~Pueschel$^{15}$,
J.~Quinn$^{9}$,
K.~Ragan$^{16}$,
P.~T.~Reynolds$^{24}$,
D.~Ribeiro$^{1}$,
E.~Roache$^{3}$,
J.~L.~Ryan$^{22}$,
I.~Sadeh$^{15}$,
M.~Santander$^{19}$,
G.~H.~Sembroski$^{25}$,
R.~Shang$^{22}$,
D.~Tak$^{15}$,
V.~V.~Vassiliev$^{22}$,
A.~Weinstein$^{7}$,
D.~A.~Williams$^{17}$,
and 
T.~J.~Williamson$^{10}$\\

\noindent
$^1${Physics Department, Columbia University, New York, NY 10027, USA}
$^{2}${Department of Physics and Astronomy, DePauw University, Greencastle, IN 46135-0037, USA}
$^3${Center for Astrophysics $|$ Harvard \& Smithsonian, Cambridge, MA 02138, USA}
$^4${Department of Physics, Washington University, St. Louis, MO 63130, USA}
$^5${Department of Physics and Astronomy, Barnard College, Columbia University, NY 10027, USA}
$^6${Physics Department, California Polytechnic State University, San Luis Obispo, CA 94307, USA} 
$^7${Department of Physics and Astronomy, Iowa State University, Ames, IA 50011, USA}
$^8${Department of Astronomy and Astrophysics, 525 Davey Lab, Pennsylvania State University, University Park, PA 16802, USA}
$^9${School of Physics, University College Dublin, Belfield, Dublin 4, Ireland}
$^{10}${Department of Physics and Astronomy and the Bartol Research Institute, University of Delaware, Newark, DE 19716, USA}
$^{11}${School of Physics and Astronomy, University of Minnesota, Minneapolis, MN 55455, USA}
$^{12}${Department of Physics, California State University - East Bay, Hayward, CA 94542, USA}
$^{13}${School of Physics and Center for Relativistic Astrophysics, Georgia Institute of Technology, 837 State Street NW, Atlanta, GA 30332-0430}
$^{14}${School of Physics, National University of Ireland Galway, University Road, Galway, Ireland}
$^{15}${DESY, Platanenallee 6, 15738 Zeuthen, Germany}
$^{16}${Physics Department, McGill University, Montreal, QC H3A 2T8, Canada}
$^{17}${Santa Cruz Institute for Particle Physics and Department of Physics, University of California, Santa Cruz, CA 95064, USA}
$^{18}${Department of Physics and Astronomy, University of Utah, Salt Lake City, UT 84112, USA}
$^{19}${Department of Physics and Astronomy, University of Alabama, Tuscaloosa, AL 35487, USA}
$^{20}${Department of Physics and Astronomy, University of Iowa, Van Allen Hall, Iowa City, IA 52242, USA}
$^{21}${Institute of Particle and Cosmos Physics, Universidad Complutense de Madrid, 28040 Madrid, Spain}
$^{22}${Department of Physics and Astronomy, University of California, Los Angeles, CA 90095, USA}
$^{23}${Institute of Physics and Astronomy, University of Potsdam, 14476 Potsdam-Golm, Germany}
$^{24}${Department of Physical Sciences, Munster Technological University, Bishopstown, Cork, T12 P928, Ireland}
$^{25}${Department of Physics and Astronomy, Purdue University, West Lafayette, IN 47907, USA}

\end{document}